\documentclass{aa}
\usepackage{psfig}
\begin{document}

\newcommand{\beq}{\begin{equation}}
\newcommand{\eeq}{\end{equation}}
\newcommand{\diff}{{\rm d}}
\newcommand{\lp}{ \left(}
\newcommand{\rp}{ \right)}
\newcommand{\cf}{{\it cf.}~}
\newcommand{\ie}{{\it i.e.}~}
\newcommand{\eg}{{\it e.g.}~}

\bibliographystyle{plain}

\title{Angular Momentum Transport by Internal Gravity Waves}
\subtitle{II - Pop II stars from the Li plateau to the horizontal branch}

\author{Suzanne Talon\inst{1} and Corinne Charbonnel\inst{2,3}}

\offprints{Suzanne Talon}

\institute{
D\'epartement de Physique, Universit\'e de Montr\'eal, Montr\'eal PQ H3C 3J7, Canada
\and Laboratoire d'Astrophysique de l'Observatoire Midi-Pyr\'en\'ees, CNRS UMR 5572, 
14, Av. E.Belin, 31400 Toulouse, France 
\and Observatoire de Gen\`eve, 51, ch. des Maillettes, 1290 Sauverny, Switzerland \\
(Suzanne.Talon@astro.umontreal.ca, Corinne.Charbonnel@obs.unige.ch)}

\date{Received / Accepted }

\authorrunning{S. Talon \& C. Charbonnel}
\titlerunning{Internal gravity waves, Li plateau and HB stars}

\abstract{
This paper is the second in a series where we examine 
the generation and filtering of internal gravity waves in stars and 
the consequences of wave induced transport of angular momentum 
at various stages of the stellar evolution. 
Here we concentrate on Pop II dwarf stars and we focus in particular on 
the differential properties of internal gravity waves as a function
of the stellar mass.
We show that, for the range of masses corresponding to the lithium
plateau, gravity waves are fully efficient and should thus lead to
a quasi-solid rotation state, similar to that of the Sun.
In the slightly more massive progenitors of currently observed horizontal
branch star however, internal wave generation is not efficient on the
main sequence, and large internal differential rotation can thus be maintained.
This leads to a natural explanation of the large rotational velocities
measured on the horizontal branch in some globular clusters.
}

\maketitle

\section{The Li plateau}
The study of lithium abundances in metal-poor halo stars has 
crucial implications for stellar structure, galactic evolution, and cosmology.
Since its discovery by Spite \& Spite (1982), the so-called lithium 
plateau\footnote{This denomination refers to the remarkably constant and flat 
lithium abundances among the metal-poor galactic halo dwarfs 
with effective temperature between $\sim$ 5700 and 6300K.}
has been the subject of considerable observational 
and theoretical effort.
Of main importance is the question of the ``uniformity'' of lithium abundances 
in Population II stars, 
and there has been a spirited debate about the existence and magnitude 
in halo star data of dispersion and trends with metallicity and 
effective temperature. (We refer to Primas \& Charbonnel
2004 for a detailed review of the literature).

On one hand, the slope with [Fe/H] mainly constrains
the contribution of galactic chemical evolution 
to the observed lithium abundance 
and in particular the rate of cosmic ray spallations in the early Galaxy 
while the trend with $T_{\rm eff}$ principally constrains
the mass dependence of the internal physics of stars.
On the other hand the existence or absence of a detectable 
range in abundance at fixed $T_{\rm eff}$ and [Fe/H] 
should be attributed either to variations in the spallation rate or
to some stellar property (\eg initial rotation rate or
magnetic field)
that affects the physical processes responsible for 
variations of the 
surface lithium abundances.

These questions are of prime importance when it comes to determine the 
primordial lithium abundance, and they recently became even more crucial 
in view of the results of the 
{\sl Wilkinson Microwave Anisotropy Probe} (WMAP; Bennett et al. 2003; Spergel et al. 2003) 
on Cosmic Microwave Background (CMB) anisotropies. 
Indeed the CMB-based determination of the baryon-to-photon ratio, $\eta$, 
corresponds to a BBN-predicted primordial lithium abundance much larger 
(by a factor $\sim$ 2.7 to 3.5) than the lithium abundances observed in the 
most metal-poor halo stars (Cyburt et al. 2003; Cuoco et al. 2003; 
Romano et al. 2003; see Charbonnel \& Primas 
2004 for more details). 
As we shall discuss below in details, 
this is crucial for stellar physics as it gives strong constraints 
on the physical mechanism(s) which is(are) 
responsible for surface lithium depletion in these objects. 

\section{Current status of Pop II stellar models}
When the first models of Pop II dwarf 
stars including atomic diffusion 
have been computed (Michaud et al. 1984), 
it became clear that slow mixing processes are at act in the radiative zones 
of these stars. 
Indeed, under conditions applicable for halo stars, pure atomic diffusion
causes He, Li and heavier elements to sink with respect to H. 
As the settling time-scale decreases with density, it is much shorter 
(at a given [Fe/H]) in the more massive stars of the plateau which 
have thinner convective envelopes.
As a result, theoretical models including pure element segregation predict 
that the degree of surface Li depletion increases with $T_{\rm eff}$. 
This result has been confirmed several times (Deliyannis et al. 1990;
Proffitt \& Michaud 1991; Chaboyer \& Demarque 1994;
Vauclair \& Charbonnel 1995; Salaris \& Weiss 2001); the case has been 
made even stronger recently with a generation of
more sophisticated models for Pop II stars
that include gravitational settling, thermal diffusion and radiative 
accelerations (Richard et al. 2002). 
The negative slope of Li with effective temperature obtained in these models 
is in contradiction with the Li data and calls for some macroscopic process 
that moderates atomic diffusion. Consistently, such a process is also required 
to counteract the settling of heavier elements in order to explain 
the close similarity of iron abundances in near turn off, sub-giant 
and lower RGB stars in globular clusters, and to reproduce the 
observed morphologies of globular cluster color-magnitude 
diagrams (see VandenBerg et al. 2002).

This is the recognition that models for Pop II stars should be
no less sophisticated than current solar models which are strongly 
constrained by helioseismology. Those indeed require both atomic diffusion
and slow mixing in the radiative zone to fulfill modern observational constraints 
(e.g., Christensen-Dalsgaard et al. 1993; Bahcall et al. 1995; Richard et al. 1996; 
Basu 1997; Brun et al. 1999).

Let us note that Salaris \& Weiss (2001) proposed an alternative solution 
to reconcile the pure (i.e., uninhibited) atomic diffusion models with the halo data. 
They showed that there is no apparent contradiction between theoretical 
lithium predictions and observations, provided that the most metal-poor field stars 
are more than 13.5 Gyr old. In that case, the more massive stars which are theoretically 
responsible for the negative lithium trend on the hot side of the
plateau would have reached the sub-giant branch; 
there, they dredge-up the lithium that has settled below their convective envelope 
during the main sequence lifetime, and their surface abundance is 
temporarily close to that 
of less massive stars still on the main sequence. 
For lower ages however the discrepancy discussed above of course remains. 
As the authors state themselves, this is not the proof that diffusion 
is fully efficient in halo stars. This means only that pure diffusion models 
can not be definitively discarded in view of the lithium halo data alone.
However the identical values of [Fe/H] 
exhibited by turnoff and more evolved stars in 
globular clusters like NGC 6397 or 47Tuc remains an objection to this
statement 
(Minniti et al. 1993; Castilho et al. 2000; Gratton et al. 2001;  
Th\'evenin et al. 2001; Carreta et al. 2003). 

Among the physical processes that have been invoked to counteract atomic diffusion
in stellar interiors, one of the most natural is rotation as already suggested in 
a more general context by Eddington (1929). 
Rotating models for Pop II stars, including or not atomic diffusion, 
have been computed by several groups using different prescriptions 
for the distribution of initial angular momenta, for the loss of angular momentum 
from a magnetic wind, and even more importantly for the evolution
of the angular momentum distribution and the resulting chemical transport 
(Vauclair 1988; Chaboyer \& Demarque 1994; Pinsonneault et al. 1991, 1999, 2002; 
Charbonnel \& Vauclair 1995; Th\'eado \& Vauclair 2001). 
Since we have no information on the history of the lithium abundances and of 
rotational velocities
(see \S 6) of halo stars, these prescriptions are generally calibrated  from
observational constraints based on open cluster stars and on the Sun. 

Classical rotating models predict that a range of initial angular momenta 
generates a range of lithium depletion and that the scatter increases 
with the average lithium depletion. For Pop II stars this 
leads to a predicted dispersion in lithium abundances along the plateau which is 
higher than observed (Chaboyer \& Demarque 1994; Vauclair \& Charbonnel 1995; 
Bonifacio \& Molaro 1997; Ryan et al. 1999). 
Pinsonneault et al. (1999, 2002) argue that this difficulty can be alleviated 
when the distribution of initial angular momenta is inferred from stellar rotation 
data in young open clusters. 
Considering the fact that the majority of young stars have low and similar 
rotation rates, they obtain a lithium depletion for halo stars of $0.1 - 0.2$ dex,  
this small value being a consequence of the observational constraint on the narrowness 
of the plateau. In view of the 
determination of $\eta$ coming from WMAP, this 
theoretical depletion factor cannot
reconciliate the data with the BBN-predicted primordial lithium abundance.

Vauclair (1999) investigated the interaction between meridional circulation
and helium settling in slowly, solid-body rotating stars 
and suggested that a quasi-equilibrium
state could be reached in which both processes would, on average,
cancel each other.
Because lithium diffuses at the same rate as helium, lithium settling would 
also be canceled by this process.
Vauclair \& Th\'eado (2001) presented numerical computations based on these developments.
For the case of slowly rotating halo stars 
they predict a lithium depletion slightly smaller than a factor two with a dispersion of 0.1 dex 
in the middle part of the plateau (the dispersion becoming larger at the two extremes 
of the plateau in effective temperature).  
These results have been obtained under the hypothesis that the stars have always 
rotated slowly (constant rotation velocity along the stellar lifetime 
with low values between 2.5 and 7.5 km.sec$^{-1}$), 
and neglecting any rotational braking in the early stages of evolution. 
As the authors say, this assumption is impossible to check directly, but it is in 
conflict with the observed spin-down of open cluster stars. 
However there is no theoretical reason why Pop I and Pop II stars should behave 
so differently. 
In the case of an important angular momentum loss, Vauclair \& Th\'eado 
expect a larger lithium destruction before their self-regulating process 
could take place, resulting probably also into a larger final dispersion. 

In summary, all current stellar models have difficulties to reconciliate 
a non negligible depletion of lithium (as needed from WMAP) with both the flatness and the 
extremely small dispersion of the lithium abundance along the plateau even when 
very special assumptions are made. 
Let us also recall that current rotating models which include only hydrodynamical 
mechanisms like meridional circulation and turbulence to transport angular momentum all share 
the same lapse : When applied to Pop I stars, they predict a substantial internal differential 
rotation for the Sun which is not compatible with helioseismology data (Libbrecht \& Morrow 1991;
Chaboyer et al. 1995; Tomczyk et al. 1995; Krishnamurthi et al. 1997; Matias \& Zahn 1998). 
This provides evidence for angular momentum transport from a mechanism which 
has still been neglected.

\section{Rotation-induced mixing and angular momentum transport in low-mass Pop I stars}
The main uncertainty in modern rotating models rests on the lack of information 
on the internal differential rotation. 
In stellar interiors indeed, the degree of rotational mixing depends on several factors, 
among which the evolution of the internal distribution of angular momentum 
is the most important.
As discussed in more details in Talon \& Charbonnel (1998, 2003a - hereafter Paper I), 
the strongest constraints on the physics of 
angular momentum transport in low-mass stars are the helioseismic 
rotation profile and the shape of the so-called Li dip. 
These independent clues point towards a mechanism that transports 
angular momentum much more efficiently than 
meridional circulation and turbulence in cool main sequence stars which 
have extended convective envelopes.
In view of the importance of the evaluation of the primordial lithium abundance, 
it is worth checking whether this process could be at act in Population II stars. 
This is one of the main goals of the present paper.

In Paper I we discussed the efficiency of angular momentum extraction  
by internal gravity waves in main sequence low-mass stars of Population I.
We showed that these waves, which are able to shape the 
solar rotation profile (Talon et al. 2002) present a peculiar mass dependence. 
We found indeed that the total momentum luminosity in waves rises with stellar mass, 
up to the quasi-disappearance of the stellar convective envelope around 6500~K
(corresponding to a mass of $\sim$1.4~$M_{\odot}$ for solar metallicity)
where the momentum luminosity drastically drops (see Fig.~3).
As a result, the net momentum extraction associated with 
internal gravity waves presents a mass dependence that explains the Li dip 
in terms of rotational mixing : 
On the hot side of the Li dip and in more massive stars, the
transport of angular momentum and of chemicals by
meridional circulation and shear instabilities do explain the lithium pattern 
(as well as the abundance anomalies of He and CNO; see Talon \& Charbonnel 1998;
Charbonnel \& Talon 1999; Palacios et al. 2003); these stars should present internal
differential rotation. 
In lower mass stars, gravity waves dominate the transport of
angular momentum,
thereby reducing the magnitude of meridional circulation and shears
and shaping the Li pattern on the cold side of the dip.
We thus predicted that Pop I main sequence stars
with initial masses lower than $\sim$1.4~$M_{\odot}$ must be quasi-solid
body rotators, as the Sun is.
Including this effect in addition to the transport of angular momentum by 
meridional circulation and turbulence is absolutely necessary in order 
to modelize correctly the rotation-induced mixing for stars on the cool side of the Li dip 
(Charbonnel \& Talon 2004).
 
In this paper we present the results of a similar study for Pop~II stars.
Again, we focus on the differential properties of internal gravity waves among 
main sequence stars of various masses. 
We also discuss 
the expected consequences on the rotation of horizontal branch stars.
Complete models including both the effects of rotation and of the 
waves will be presented in a forthcoming paper. 

\section{Internal gravity waves in Pop II stars along the Li plateau}

\subsection{Input physics}
We refer to Paper~I for details on internal waves physics and on the computational method,
and we just summarize here the key points.

\begin{enumerate}
\item {\em Excitation mechanism}. 
In order to evaluate the wave spectrum produced 
by stellar convective envelopes, we consider 
the excitation associated with
Reynolds stresses as described by Goldreich et al. (1994)
(see \S~2.2, Paper~I).
Excitation by overshooting is neglected due to the lack of a reliable
prescription.
Note that we expect overshooting to produce mostly high-$\ell$
waves which are not efficient in momentum extraction from the deep
interior. This point is discussed further in Paper~I.
\item {\em Shear layer}. The wave-mean flow interaction leads to the formation
of a thin, double peaked shear layer,
that oscillates on a short time-scale (of the order of years). 
This layer, whose amplitude is self regulated by shear turbulence 
(this point will be discussed in details in Talon \& Charbonnel 2004), 
acts as a filter on the low $\ell$ waves that carry angular momentum
in the inner radiative zone (see \S~3, Paper~I).
\item {\em Momentum extraction}. We follow the time evolution of the shear layer according 
to Eq.(7) of Paper I over several oscillation cycles. 
In the presence of differential rotation between the convection zone
and the bottom of the shear layer, 
differential wave dissipation between low prograde and retrograde waves
leads to a net momentum deposition below this layer
which varies with differential rotation (see \S 4.1, Paper I). 
We calculate the net, average luminosity below the filtering shear layer 
(at $r=R_{\rm cz}-0.03~R_*$) for 
a given value of the differential rotation.
It corresponds to momentum extraction in the stellar interior
when the core rotates faster than the surface. 
\end{enumerate}

We examine the properties of internal gravity waves in stars with masses 
between 0.6 and 0.9~$M_{\odot}$ and with $[{\rm Fe/H}] = -2$.
We consider a value of 0.2431 for the initial He mass fraction 
and the relative concentrations of $\alpha$-elements are increased relative to the 
solar mix as appropriate for Pop~II stars. 
We use the same code and input physics (eos, opacities, nuclear reactions) 
as in Palacios et al. (2002). 
Table~1 summarizes the model characteristics that are 
relevant for our purpose. 

\begin{table*}
\caption{Characteristics of the stellar models (with $[{\rm Fe/H}]=-2$)
on the ZAMS and at 10~Gyr :
Effective temperature, luminosity,
thickness of the external convective zone in stellar mass and radius, 
largest convective scale $\ell_c$, characteristic convective frequency $\nu_c$, 
and thermal diffusivity below the convective envelope $K_T$. 
$\nu_{\rm max}$ is the maximum frequency required
to capture the wave - mean flow interaction.}
\begin{center}
\begin{tabular}{ccccccccccc}
\hline
\multicolumn{1}{c}{$M_*$} & \multicolumn{1}{c}{Age}
& \multicolumn{1}{c}{$T_{\rm eff}$}
& \multicolumn{1}{c}{$L/L_{\odot}$}
& \multicolumn{1}{c}{log$\left({M_{\rm cz}/{M_*}}\right)$}
& \multicolumn{1}{c}{$R_{\rm cz}/{R_*}$}
& \multicolumn{1}{c}{$\ell_c$} 
& \multicolumn{1}{c}{$\nu_c$}
& \multicolumn{1}{c}{$K_T$}
& \multicolumn{1}{c}{$\nu_{\rm max}$}
\\
($M_{\odot}$) & (Gyr) & (K) & & & & &($\mu$Hz) &
(cm$^2$~s$^{-1}$) & ($\mu$Hz)\\
\hline
0.60 & zams & 4985 & 0.14 & $-1.18$ & 0.271 & 57 & 0.517 & $2.43 \times 10^5$ & 1.25 \\
     & 10   & 5125 & 0.18 & $-1.30$ & 0.263 & 59 & 0.614 & $4.38 \times 10^5$ & 1.25\vspace{0.1cm}\\
0.65 & zams & 5245 & 0.21 & $-1.44$ & 0.236 & 67 & 0.751 & $7.79 \times 10^5$ & 1.25 \\
     & 10   & 5436 & 0.29 & $-1.66$ & 0.229 & 69 & 0.879 & $1.71 \times 10^6$ & 1.25\vspace{0.1cm}\\
0.70 & zams & 5490 & 0.30 & $-1.77$ & 0.206 & 78 & 1.05 & $2.37 \times 10^6$ & 1.25 \\
     & 10   & 5725 & 0.47 & $-2.22$ & 0.181 & 90 & 1.68 & $1.14 \times 10^7$ & 1.25\vspace{0.1cm}\\
0.75 & zams & 5720 & 0.42 & $-2.18$ & 0.176 & 92 & 1.50 & $6.89 \times 10^6$ & 1.25 \\
     & 10   & 6015 & 0.78 & $-3.15$ & 0.125 & 132& 4.27 & $1.68 \times 10^8$ & 1.85\vspace{0.1cm}\\
0.80 & zams & 5935 & 0.57 & $-2.79$ & 0.131 & 126 & 3.29 & $4.31 \times 10^7$ & 1.85 \\
     & 10   & 6342 & 1.37 & $-4.15$ & 0.065 & 259 & 26.1 & $1.42 \times 10^{10}$ & 6.25\vspace{0.1cm}\\
0.82 & zams & 6010 & 0.64 & $-3.04$ & 0.119 & 140 & 3.96 & $6.56 \times 10^7$ & 1.85 \\
     & 10   & 6503 & 1.80 & $-5.96$ & 0.041 & 392 & 8.02 & $2.91 \times 10^{11}$ & 17.5\vspace{0.1cm}\\
0.85 & zams & 6140 & 0.76 & $-3.55$ & 0.094 & 175 & 7.02 & $2.22 \times 10^8$ & 2.50 \\
     & 10   & 6810 & 2.92 & $-8.23$ & 0.010 & 1110& 384  & $2.80 \times 10^{13}$ & $*$\vspace{0.1cm}\\
0.9  & zams & 6370 & 0.99 & $-4.56$ & 0.062 &270 & 20.9 & $2.39 \times 10^9$ & 3.75 \\
\hline
\multicolumn{10}{l}{$*$ {\small No oscillation occurs in this model.}}
\end{tabular}
\end{center}
\end{table*}

\subsection{Momentum extraction by internal gravity waves along the Li plateau}
The generation of internal gravity waves depends on the structure of the 
stellar convective envelope. 
For the stars we consider here this region is relatively extended
as can be seen on Fig.~1 (see also Table~1) where some of their
characteristics are plotted as a function of effective temperature both 
on the zero age main sequence (zams; open squares) and 
at 10~Gyr (black squares).
At a given age, the size of the convective envelope 
decreases when one goes to higher effective temperature. 
For a given stellar mass, 
as the star evolves the effective temperature rises and thus the thickness 
of the convective envelope decreases.

Let us note that not all the stars for which we made the 
present computations will lie on the lithium plateau itself. 
If we consider indeed an age of 10 Gyr, only the stars with masses between 
$\sim 0.7~{\rm M}_{\odot}$ and $\sim 0.82~{\rm M}_{\odot}$ (for $[{\rm Fe/H}]=-2$) 
have an effective temperature in the plateau range (i.e., between $\sim$ 5700 and 6500K),
while cooler ones correspond to stars that exhibit lower lithium abundances and 
more massive ones have already evolved off the main sequence. 
In this limited mass range (which is indicated by the shaded area in Fig.~3 where 
we show the corresponding range in effective temperature on the zams), 
the thickness of the convective envelope varies only modestly : 
from $\sim 20 \%$ to $\sim 12 \%$ of the stellar radius for the zams models. 
In the same mass range and still on the zams, the characteristic convective frequency 
varies from $\nu_c \sim$1.6~$\mu$Hz to $\sim$12~$\mu$Hz while the spherical
harmonic number $\ell_c$ corresponding to the largest convective scale varies from 
$\sim$80 to $\sim$140.

In order to investigate how the wave spectrum produced by these convection zones varies 
as a function of stellar mass, 
let us first discuss waves characteristics based on the 
stellar models on the zero age main sequence.
The corresponding momentum spectra are shown in Fig.~2 
where ${\cal L}_J=4\pi r_{\rm cz}^2 {\cal F}_J$ is the momentum luminosity, 
$r_{\rm cz}$ and ${\cal F}_J$ being respectively the radius at the base of the convective 
envelope and the momentum flux. 
We see that the spectrum characteristics, which depend on 
the structure of the convection zone, evolve with stellar mass. 
In particular, low frequency waves disappear when the stellar mass increases due to 
a larger thermal diffusivity below the thinner convective envelope (see Fig.~3) 
that leads to stronger damping (Eq.~1). In addition, the flux associated 
with a given frequency also rises with stellar mass together with the convective flux.

We then calculate the net momentum extraction associated 
with the waves for an initial 
differential rotation of $\delta \Omega = 0.01 \,\mu {\rm Hz}$ over 
$0.05\,R_*$\footnote{This differential rotation is somewhat smaller than
the value used in Paper~I. In the figures, points corresponding to
Pop~I stars are calculated using the same model as those in
Paper~I, but with the new preferred value for the differential
rotation.}. 
Figure~3 presents the corresponding net average luminosity at $0.03\,R_*$
below the surface convection zone as a function of $T_{\rm eff}$. 
We see that on the zams the net momentum luminosity modestly increases with the 
stellar mass (and thus with $T_{\rm eff}$) up to $\sim 0.8~{\rm M}_{\odot}$ 
and then slightly decreases for more massive stars before dropping dramatically 
in the $\sim 0.9~M_{\odot}$ star.
For the limited mass range corresponding to the lithium plateau, the net momentum 
luminosity actually presents also a plateau. 

During the stellar life, the characteristics of the convective envelope evolve 
(see above), and consequently the net momentum luminosity changes with respect 
to its zams value.
Of main importance are the changes of the thermal diffusivity below the 
convective envelope. 
At 10 Gyr, this quantity has indeed considerably increased for stars 
with masses higher than $0.8~M_{\odot}$ resulting in a strong decrease of 
the filtered momentum luminosity in these objects with respect to its zams value.
However for the cooler stars which are less evolved at the same age, $K_T$ remains closer
to its initial value.
Very importantly, we have to note that 
the dependence of the net momentum luminosity
as a function of effective temperature is very similar on the zams and at 10 Gyr :
It presents a plateau as do the lithium abundances. 

\begin{figure}[t]
\centerline {
\psfig{figure=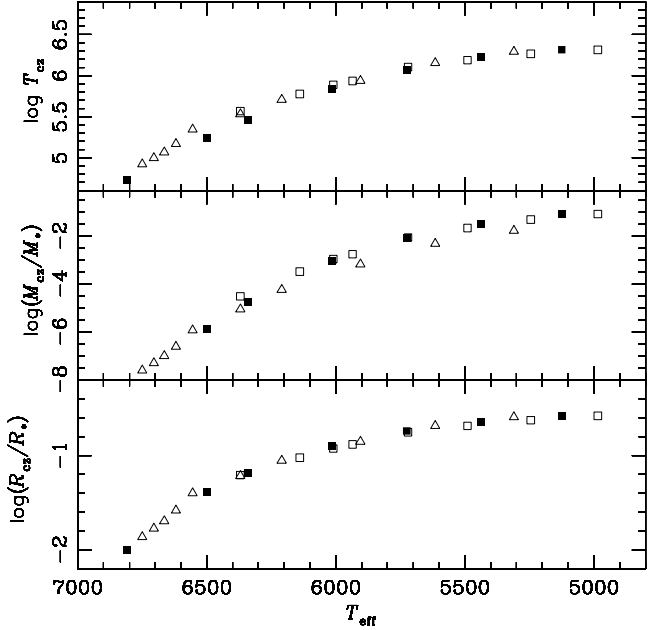,height=9cm,angle=0}
}
\caption{({\em top}) Temperature at the base of the convection zone ($T_{\rm cz}$), 
({\em middle}) mass of the convection zone 
and ({\em bottom}) radius of the convection zone 
as a function of $T_{\rm eff}$. 
Squares : Pop~II stars on the zams (open squares) and at 10 Gyr (black squares) 
(masses are those listed in Table~1); 
Triangles: Pop~I stars on the zams
(models from Paper I). 
\label{basezc2}}
\end{figure}

\begin{figure}[t]
\centerline {
\psfig{figure=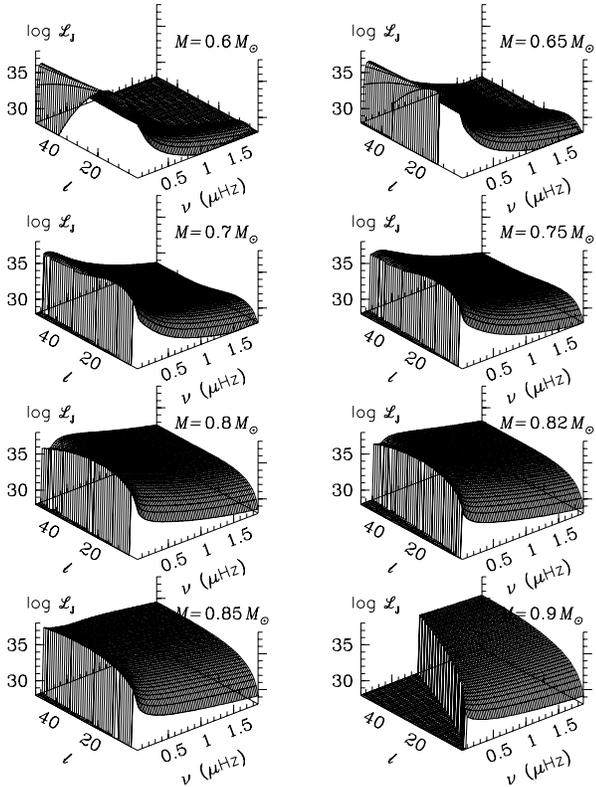,height=12cm}
}
\caption{Momentum wave spectrum in Pop~II stars 
of various masses on the zams
for the excitation model by Goldreich et al. (1994). 
The small radiative conductivity allows very low frequency waves 
to contribute to momentum transport.
\label{spectre}}
\end{figure}

\begin{figure}[t]
\centerline {
\psfig{figure=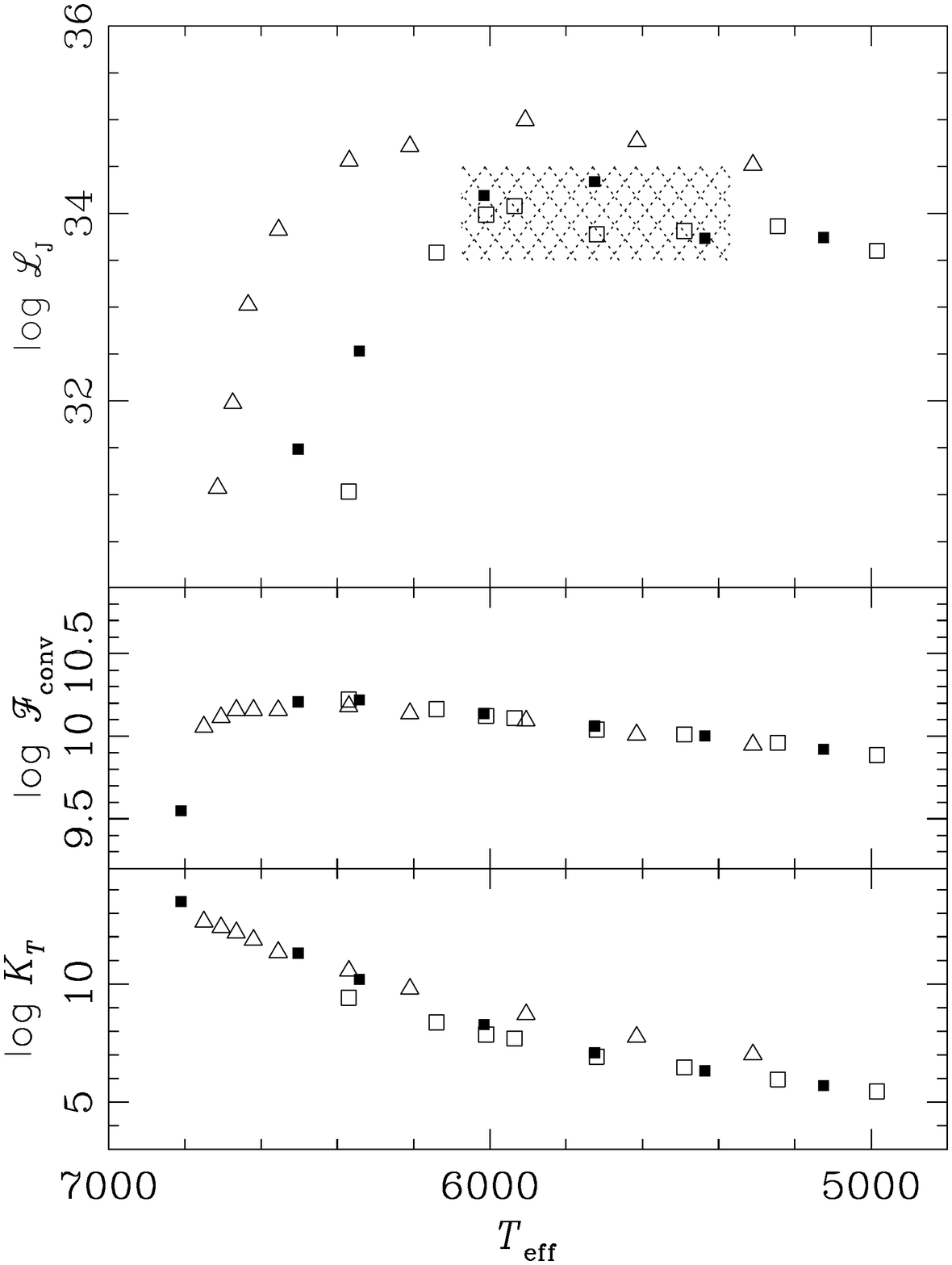,height=12cm}
}
\caption{({\em top}) Net momentum luminosity at $0.03\,R_*$ 
below the surface convection zone as a function of $T_{\rm eff}$ 
for an initial differential rotation of 
$\delta \Omega = 0.01\,\mu {\rm Hz}$ over $0.05\,R_*$. 
({\em middle}) Convective flux as a function of $T_{\rm eff}$.
({\em bottom}) Thermal diffusivity as a function of $T_{\rm eff}$.
Symbols are defined in Fig.~1.
The shaded area indicates the range in zams effective temperature of 
the Pop~II stars which will lie on the lithium plateau at 10~Gyr.  
\label{basezc2}}
\end{figure}

\subsection{Comparison with Pop I stars on the cool side of the Li dip}
Before comparing wave characteristics in Pop~I and Pop~II stars, let us 
underline some important points concerning the stellar models.
At a given $T_{\rm eff}$, Pop~II stars along the Li plateau have lower masses than 
Pop~I stars on the cool side of the Li dip. 
Regarding the depth of convective envelopes however, their lower metallicity 
tends to compensate the mass effect. This can be seen on Fig.~1 where the thickness 
in both mass and radius of the convective envelope and the temperature at its base 
are plotted for both sets of models as a function of $T_{\rm eff}$ 
(triangles correspond to the Pop~I zams models of Paper I).
On the zams, these characteristics are extremely close for both stellar populations. 
As a result, the convective fluxes (${\cal F}_{\rm conv}$) are almost identical 
in both populations (Fig.~3).
Let us note however that, for a given effective temperature, the thermal 
diffusivity ($K_T$) below the convective envelope 
is always lower in Pop II stars than in their more metal-rich counterparts. 
Thermal diffusivity plays a major role in determining which frequencies
contribute to momentum deposition in the interior. Indeed, the local wave
amplitude is proportional to
$\exp \left[ -\tau(r, \sigma, \ell)\right]$ with
\begin{eqnarray}
\lefteqn{\tau(r, \sigma, \ell) = }  \label{optdepth} \\
&&[\ell(\ell+1)]^{3\over2} \int_r^{r_c} 
\lp K_T + \nu_t \rp \; {N N_T^2 \over
\sigma^4}  \left({N^2 \over N^2 - \sigma^2}\right)^{1 \over 2} {\diff r
\over r^3} \nonumber
\end{eqnarray}
where $\sigma$ is the local, Doppler shifted frequency,
$N^2 = N_T^2 + N_{\mu}^2$ is the Brunt-V\"ais\"al\"a frequency, 
$N_T^2$ is its thermal part 
and $ N_{\mu}^2$ is due to the mean molecular weight stratification
(see Zahn et al. 1997 for details).
Differential filtering occurs if the ratio $K_T/\sigma^4$ leads to 
significant differential damping
between prograde and retrograde waves\footnote{This will be the case if
differential rotation is of the same order as $\omega$ and $K_T$ is large
enough.}, while a large thermal diffusivity
totally inhibits low frequency waves. Thus the lower the thermal diffusivity,
the lower are the low cut-off frequency (see Fig.~2) and the
maximum frequency that leads to differential filtering 
($\nu_{\rm max}$)\footnote{$\nu_{\rm max}$, the maximum frequency required 
to capture the 
wave-mean flow interaction, is determined empirically.}
(see Table~1). 

Comparing the wave spectra to those obtained in Paper~I,
we find that 
the lower radiative conductivity in Pop~II models allows waves of much lower 
frequency to contribute to momentum 
redistribution.
In addition the lower thermal diffusivity below the convective envelope 
along the plateau mass range implies lower damping and the remaining of low frequency waves.

Of main importance for the differential effect of the waves from star to star
is the increase of thermal diffusivity at higher $T_{\rm eff}$. This translates into 
increasing radiative damping which becomes critical for $T_{\rm eff}$ higher than 
$\sim 6200$~K. 
The net momentum luminosity is shown for both populations in Fig.~3. 
Its general behavior as a function of effective temperature is very similar, 
while its magnitude is lower for Pop~II stars.
This is a consequence of the smaller radius of Pop~II stars.

\begin{figure}[t]
\centerline {
\psfig{figure=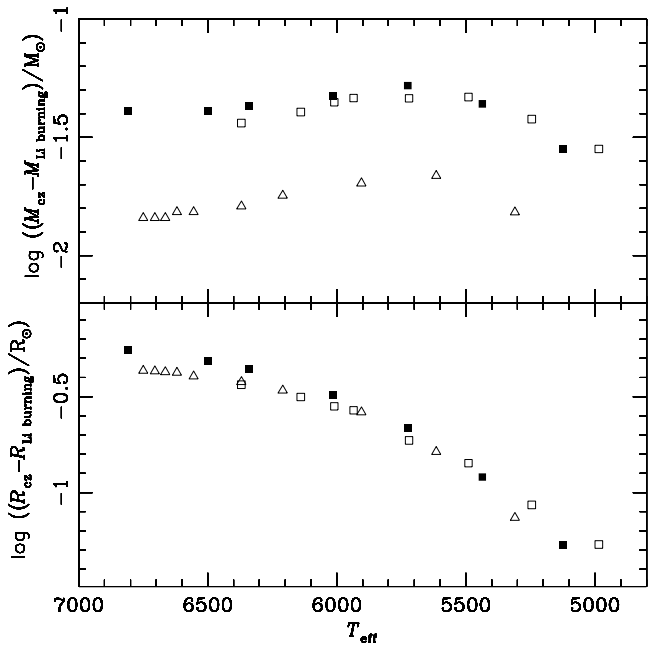,height=9cm}
}
\caption{Measure of the ``distance'' between the lithium burning
region and the base of the convection zone with respect to
$T_{\rm eff}$. ({\em top}) In mass units.
({\em bottom}) In radius units.
Symbols are defined in Fig.~1.
\label{basezc4}}
\end{figure}

\section{Consequences for the Li depletion in Pop~II stars}
\subsection{From a coherent picture of mixing in Pop~I stars ...}
Combining the shape of the lithium dip and the dependence
with stellar mass of the net wave momentum luminosity,
we could conclude in Paper I that internal gravity waves are able to 
enforce solid-body rotation for Pop~I stars with initial masses 
lower than $\sim 1.4~M_{\odot}$, i.e., with effective temperatures 
lower than $\sim 6500$~K. 
There, the differential rotation (that determines the extent
of lithium destruction) will be the result of an
equilibrium between momentum transport by meridional circulation,
turbulence and internal gravity waves.
 
The time required for a star of a given mass to rotate as a solid body 
remains to be computed, but the solar case gives some clues.
Helioseismology indeed constrains the complete 
rotation profile of a $1~M_\odot$ Pop~I star, at 4.5~Gyr.
This is however only an upper limit to the time-scale required to enforce solid body 
rotation in a solar-type star. 
Using the same physics as we presently do, 
Talon et al. (2002) have shown that internal gravity waves 
can establish an almost uniform profile in a solar-like star in about 
10$^7$~yr. 

For the purpose of calculating lithium evolution, 
what is actually important is not the whole star, but rather the external region
between the base of the convective envelope and the region of lithium destruction.
Below the surface shear layer (i.e., just below the convective region),
internal gravity waves will contribute to reduce differential rotation,
thus limiting lithium destruction. 
Lithium abundances in open clusters give constraints on   
the time evolution of the rotation profile in these outer radiative regions. 
As can be seen from the lithium data in young clusters, 
this time-scale is extremely short for stars on the cold side
of the lithium dip for which the net momentum luminosity reaches 
important values. In a very young cluster like the Pleiades (age $\sim 0.135$~Gyr
according to the WEBDA, Mermilliod\footnote{http:$//$obswww.unige.ch$/$webda}), 
the lithium dip, though not very pronounced, begins to appear 
(i.e., King et al. 2000). 
This feature is very clear in Coma Ber (i.e., Ford et al. 2001), 
which is $\sim$ 0.45 Gyr (WEBDA, Mermilliod). 
At this age, stars with effective temperatures higher than $\sim 6500$~K 
already show very strong lithium depletion, unlike their slightly cooler counterparts 
which have undergone only modest depletion
despite their stronger braking. 
This gives an upper limit,
which is for sure highly conservative,
to the time-scale for the internal gravity waves to
balance the transport by meridional circulation and shear turbulence
in the external parts of the radiative 
region; this time-scale should be used to constrain the wave generation
model.

\subsection{... to a prescription for Pop~II stars}
In order to extrapolate this estimate to the case of Pop~II stars, 
we have to introduce an important quantity.
That is, the ratio of the stellar luminosity to the stellar mass, 
which is shown in Fig.~5 for both Pop~I and Pop~II stars.
Indeed, the time-scale for meridional circulation and turbulence 
depends on $L_*/M_*$.
This ratio strongly increases with the stellar mass
leading to more efficient rotation-induced mixing 
in more massive stars. 
This statement has to be moderated here by the knowledge that stars
with an extensive convective envelope are strongly spun down by
magnetic torques when surface rotational velocities are large,
as initially suggested by Schatzman (1962). The internal differential
rotation induced by the braking of the external convective region
enhances internal mixing, which will thus be strongly dependent
on the existing processes for angular momentum transport.

More importantly here, $L_*/M_*$ is much lower for the halo stars 
than for the metal-rich stars.  
This means that the characteristic time-scale for meridional circulation and turbulence 
will be longer (i.e., the associated mixing less efficient) 
in the halo stars. On the other hand, the net momentum luminosity is 
lower in these latter objects. How can both effect compensate?

Let us consider the case of two stars which lie in the region of full
efficiency of the internal gravity waves, and which have similar 
effective temperature on the zams : the 1.20 and $0.82~M_{\odot}$ models.
Their $L_*/M_*$(zams) values are respectively 1.34 and 0.78, while their 
${\cal L}_J$ values are 9.8$\times 10^{34}$ and 9.7$\times 10^{33}$. 
We can thus very crudely estimate that the time-scale for the
gravity waves to dominate over the hydrodynamical processes in the region
between the base of the convective envelope and that of lithium destruction
in the metal-poor star will only be only a factor five longer than that necessary in
the case of the metal-rich star. This gives an upper limit of the order 
of 2.2~Gyr if we consider the Coma Ber case. 
Let us note that this comparison is allowed by the fact that the 
distance between the base of the convective envelope and the region where Li 
burns is very similar for Pop I and Pop II stars at a given effective temperature 
(see Fig.~4). 
Again, this value is a very conservative upper limit. 

It is important to insist on the fact that the net momentum luminosity associated 
with the internal gravity waves presents a plateau in the range in 
effective temperature corresponding to that of the lithium plateau 
(In Fig.~3, the shaded area indicates the effective temperature on the zams 
of the stars which will still lie on the lithium plateau at 10 Gyr). 
To explain the shape of the plateau, several features must be taken
into account. Firstly, combining Figs~3 and 5 shows that, for
stars colder than 6100~K (the hot side of the plateau on the zams),
wave momentum luminosity and $L_*/M_*$ show the same trend,
both rising very slowly with mass. This would lead to similar mixing
efficiencies. Now adding Fig.~4, we see that, between $~$5400 and $6100$~K
(the location of the plateau on the zams),
the ``distance'' between the base of the convection zone and the location
of lithium burning is similar, while it 
starts shrinking in cooler stars.
Roughly speaking, we thus expect similar lithium destruction on
the plateau, getting more efficient for cooler stars.
Complete stellar models are required to check this conjecture, 
and will be presented in a forthcoming paper 
where the transport of both angular momentum and of the chemicals
will be computed under the effects of the meridional circulation, 
turbulence and gravity waves.

\begin{figure}[t]
\centerline {
\psfig{figure=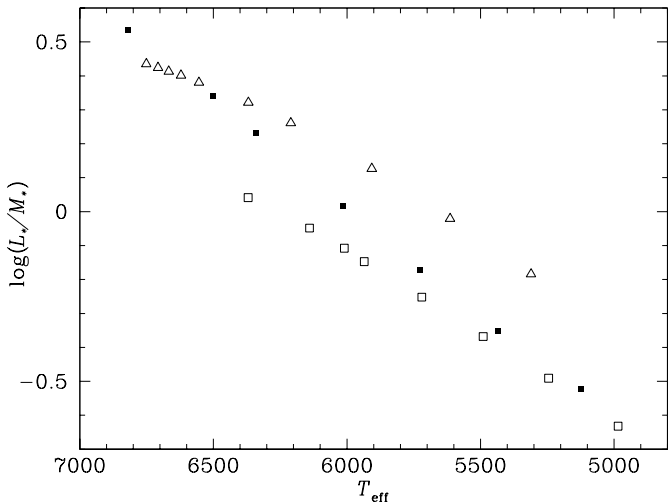,height=7cm,angle=270}
}
\caption{
Ratio of the stellar luminosity to the stellar mass as a function of
effective temperature for Pop~I and Pop~II models.
Symbols are defined in Fig.~1.
\label{lummasse}}
\end{figure}

\section{New perspectives on the rotation history of Pop~II low-mass stars}
\subsection{Surface rotation velocity of Pop~II main sequence stars}
Very few data exist on the surface rotation velocity of metal-poor dwarfs.
Rotation has been unambiguously detected only in tidally locked binaries 
(Peterson et al. 1980) or blue stragglers (Carney \& Peterson 1981). 
For single main sequence turnoff stars, 
Peterson et al. (1983) obtained an upper limit on $v~{\rm sin}i$ of 
8~km.s$^{-1}$ for 30 main sequence halo stars with metallicity 
less than 1/10 solar. 
In their very high resolution and high signal-to-noise ratio studies of the 
lithium isotope ratios in metal-poor halo stars, Hobbs et al. (1999) 
found no evidence for extra-line broadening above 4~km.sec$^{-1}$, although 
line broadening at that level was needed to explain the data.  
Finally, Lucatello \& Gratton (2003) derived upper limits to the rotation 
of turnoff and sub-giant stars in three globular clusters which are respectively 
of about 3.5 and 4.7~km.sec$^{-1}$.
Of course, we have no information on the initial rotation of these stars.
There is however no theoretical reason to believe that the distribution of 
initial momenta and the evolution of the surface rotation of low-mass Pop II stars 
have been very different from that of their Pop I counterparts.

\subsection{Rotation of horizontal branch stars}
Evolved stars can in principle provide some clues on the internal rotation
of their progenitors. In the present context, informations on the rotation 
of horizontal branch (HB) stars are precious. 
Both field and globular cluster HB stars (both blue and red) can present 
substantial rotation rates, as discovered by Peterson et al. (1983) and Peterson (1983).
For these objects $v~{\rm sin}i$ does not depend on any cluster parameter 
and neither the size nor the distribution of rotation is strongly correlated with 
the horizontal branch morphology,  
but the distribution of $v~{\rm sin}i$ varies significantly from cluster to cluster 
(Peterson et al. 1995; Cohen \& McCarthy 1997; Recio-Blanco et al. 2002; Behr 2003a). 
There is now some evidence that the hottest blue HB stars (which are supposed to have undergone 
the highest mass loss while on the red giant branch) rotate more slowly 
than their cooler counterparts (Behr et al. 2000; Recio-Blanco et al. 2002;
Behr 2003a)\footnote{In all 
the globular clusters studied up to now, all the HB stars hotter than $T_{\rm eff} \sim 11,500$~K 
have $v~{\rm sin}i \leq$ 8-12 km.sec$^{-1}$ (Recio-Blanco et al. 2002). 
Among the cooler HB stars, there is a range of rotation rates and fast rotators appear.}. 
Also, there is no apparent correlation between rotational velocity and luminosity distance
from the zero age HB. 
Field HB stars behave similarly to such stars in globular clusters 
and the red ones appear to have about the same amount of surface angular momentum
as do their blue counterparts (Preston 1997; Kinman et al. 2000; Carney et al. 2003; Behr 2003b).

In some rare cases, the observed rapid rotation rates
are well explained either by tidal locking in a binary system (Carney et al. 2003)
or by absorption of nearby planets (Soker 1998; Siess \& Livio 1999).
This is however not the case of most stars; indeed,
several considerations argue against the possibility that 
all fast rotators gained the corresponding angular momentum 
through spin-up by a companion (Peterson et al. 1995).  

\subsection{Additional clues to the mass dependence of internal gravity waves efficiency}
Another possibility is that the angular momentum was present at the time of 
stellar formation
and remained hidden in the inner regions while only the surface was spun-down
by the same process believed to operate 
in cool main-sequence Pop~I stars;
momentum then emerged at the end of the 
red giant branch when substantial mass loss occurred. 
Differential rotation with depth is a natural explanation for 
rapid rotation in HB stars to rotate rapidly while their main sequence progenitors 
where slow rotators. 
This requires a substantial differential rotation with depth while on the 
main sequence in order to preserve a large enough reservoir of internal angular momentum 
(Pinsonneault et al. 1991; Sills \& Pinsonneault 2000).

This latter interpretation has been discarded in light of the helioseismic 
determination of the internal solar rotation\footnote{In view of the 
solar data, Sills \& Pinsonneault (2000) favor the class of models 
with constant specific angular momentum (i.e., differential rotation) 
in giant branch envelopes, retention of a rapidly rotating core on the RGB, 
and subsequent angular momentum redistribution from the core to the envelope on the HB.
Even this class of models requires relatively high surface rotation 
($\sim4$~km.sec$^{-1}$) at the main sequence turnoff to explain the high rotation
rates for the cooler HB stars.}. 
It can be however reconciled both with helioseismology and with our results 
for the following reason. 
Though the present mass of the globular cluster HB stars is lower than the 
inferred turnoff mass of the corresponding globular cluster, 
their initial mass was of course higher,
implying a non negligible mass loss on the red giant branch 
($\sim 25 \%$ with a 1$\sigma$ spread of about 10$\%$ around that mean; 
Rood 1973, Lee et al. 1990).

Typically, for the very metal-deficient globular cluster M92 
([Fe/H]~=~$-2.24$ or $-2.10$ according respectively to Zinn \& West 1984 
and Carreta \& Gratton 1997, 
a value which is close to the metallicity of the present models) 
the most recent age determinations vary between 
12.3~Gyr for standard models (Salaris \& Weiss 2002, hereafter SW02) and 
13.5~Gyr for models including atomic diffusion and turbulent mixing 
(VandenBerg et al. 2002, hereafter V02). 
The corresponding turnoff mass and effective temperature are 
$\sim$ 0.78~$M{_\odot}$ and 6650~K in the SW02 models (Salaris \& Weiss, 
private communication)
and $\sim$ 0.77~$M{_\odot}$ and 6450~K in the V02 models (Richard, private communication).
Because of some differences in the input physics (in particular, SW02 and V02 assume a
lower cosmological helium mass fraction than we do, and V02 models include atomic diffusion), 
these masses can not be directly compared with the ones of our models. 
What matters however is the effective temperature of these stars while they were 
on the main sequence, and which 
is strongly connected to the properties of the convective envelope that 
determine the gravity waves efficiency. 
In both sets of models, it is higher than the critical value where 
internal gravity waves are found to fail to enforce rigid rotation.
According to the SW02 models, the turnoff effective temperature of the other globular 
clusters for which HB rotation velocities are now available (i.e., M3, M13, M15, M68, M92 
and NGC 288, Behr 2003a) is also always higher than this critical value.

This means that the stars which are at present on the HB in all the currently 
investigated globular clusters had such 
main sequence masses and effective temperatures, 
and thus such convective envelopes, that internal gravity waves
were inefficient to extract angular momentum while on the main sequence. 
Like in the more massive and more metal-rich stars which lie on the 
left side of the lithium dip in open clusters, only hydrodynamical processes 
must have been at act in their main sequence progenitors.
As a consequence the progenitors of the present HB stars must have left 
the main sequence with important internal differential rotation, 
contrary to what we predict for the stars along the lithium plateau, for the Sun 
and for the Pop I stars lying on the cool side of the lithium dip.

\section{A final remark on HB stars and transport of angular momentum}
While our main sequence predictions 
open new perspectives to explain the data on HB rotation, 
one has to remain cautious before definitively concluding on this point.
It is indeed necessary to investigate the possible effect of internal gravity 
waves while the star evolves on the red giant branch and on the HB, 
and to make complete simulations of the internal evolution of angular momentum 
all along the stellar life taking into account all the possible mechanisms 
together with a good description of the mass loss on the red giant branch. 
A self-consistent model should also explain why only a third of HB stars are fast rotators,
and how fast rotation is coupled with cool effective temperature on the HB. 
Work is in progress in this direction and will be presented in a forthcoming paper.

Let us make a final remark on the importance of HB stars rotation rates.
Up to now, two processes have been proposed to explain the helioseismic
solar rotation profile: 

{\bf 1)} In the gravity wave model (Talon et al. 2002), there is a range
of masses for which waves are efficient enough to lead to quasi-solid body
rotation. In that framework, the lithium gap is explained by the
combined effect of braking (which leads to increased differential rotation and rotational
mixing) and the appearance of waves as the surface convection zone
deepens for smaller masses (Paper~I). For main sequence stars with only shallow surface convection
zones, no waves are generated and large internal differential
rotation is expected (see \eg Talon et al. 1997, Meynet \& Maeder 2000,
Palacios et al. 2003), naturally explaining the fast rotation of HB stars.

{\bf 2)} In the magnetic model (Charbonneau \& MacGregor 1993), 
there is no relation between the presence of a surface
convection zone and solid body rotation as the dipolar field of
the radiative zone has to be of fossil origin. In that case, the lithium gap
might be explained in purely solid body rotation by the combined action
of radiative forces and solid body meridional
circulation (Charbonneau \& Michaud 1988). However, one would then
expect all the 
HB stars, originating from main sequence stars in solid
body rotation, to rotate more slowly then observed. 

All these results over a large range of the HR diagram 
strongly favor the gravity wave model. 
This does not mean of course that magnetic fields do not play a role in the
whole picture. It seems to us however that we have sufficient clues 
to push forward complete tests of a fully hydrodynamical model before 
including MHD effects.

\section{Conclusion}
In this paper, we discuss the 
generation and filtering of internal gravity waves in Pop~II dwarf stars 
and its consequences on the angular momentum distribution both on the 
main sequence and on the horizontal branch.
By comparing them with Pop~I stars, we expect gravity waves to
be efficient in producing quasi solid-body rotation in stars
of the lithium plateau on very short time-scales. 
Consequently, a new generation of stellar models must 
take into account the effects of atomic diffusion, meridional 
circulation, turbulence and gravity waves, in order to estimate 
the theoretical depletion of lithium along the plateau. 

On the other hand, we found the same dependence of 
gravity waves
efficiency with effective temperature 
as in Pop~I stars. 
In particular, gravity waves appear to be inefficient in the 
main sequence progenitors of 
present HB stars allowing 
strong differential rotation which leads 
to a natural explanation of the fast HB rotators. 

In this paper, we mainly focused on estimating whether 
gravity waves can play a significant role in shaping the internal
rotation profile of lithium plateau and horizontal branch stars and
did not intend to cover the complete problem which is far more complex.
Indeed, gravity waves can also lead to significant mixing by themselves
(see e.g. Press 1981, Schatzman 1993, Garc\`\i a L\`opez \& Spruit 1991,
Montalb\'an 1994) and proper treatment of these effects must be
taken into account when building complete stellar models with
waves. This will be discussed at length in a forthcoming paper.

Let us further note that the conjectures presented in this paper 
neglect the possible presence of a magnetic field. Indeed, our goal here
is to verify whether a purely hydrodynamical mechanism can be
responsible for the evolution of stellar parameters such as surface
velocities and abundances and to produce definite diagnoses in order to
obtain tests of our models.
Both theoretical and observational evidences show that this is 
certainly the case.

\begin{acknowledgements}
We acknowledge financial support from the 
French Programme National de Physique Stellaire (PNPS).
We thank the R\'eseau qu\'eb\'ecois de calcul de haute performance (RQCHP) 
and the Centre informatique national de l'enseignement sup\'erieur (CINES) 
for useful computational resources.
Thanks also to M.~Salaris,
A.Weiss and O.~Richard for providing informations on their
stellar models.
\end{acknowledgements}

\end{document}